**EMERGING TECHNOLOGIES**

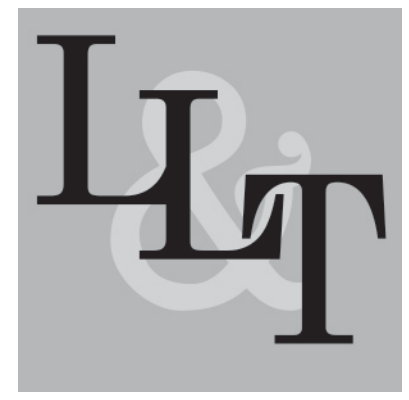

# Profiling learners' affective engagement: Emotion AI, intercultural pragmatics, and language learning

*Robert Godwin-Jones, Virginia Commonwealth University*

## Abstract

*Learning another language can be a highly emotional process, typically characterized by numerous frustrations and triumphs, big and small. For most learners, language learning does not follow a linear, predictable path, its zigzag course shaped by motivational (or demotivating) variables such as personal characteristics, teacher/peer relationships, learning materials, and dreams of a future L2 (second language) self. While some aspects of language learning (reading, grammar) are relatively mechanical, others can be stressful and unpredictable, especially conversing in the target language. That experience necessitates not only knowledge of structure and lexis, but also the ability to use the language in ways that are appropriate to the social and cultural context. A new opportunity to practice conversational abilities has arrived through the availability of AI chatbots, with both advantages (responsive, non-judgmental) and drawbacks (emotionally void, culturally biased). This column explores aspects of emotion as they arise in technology use and in particular how automatic emotion recognition and simulated human responsiveness in AI systems interface with language learning and the development of pragmatic and interactional competence. Emotion AI—the algorithmically driven interpretation of users' affective signals—has been seen as enabling greater personalized learning, adapting to perceived learner cognitive and emotional states. Others warn of emotional manipulation and inappropriate and ineffective user profiling.*

## Introduction

Given the close connection between the languages we speak and personal identity, it is no surprise that emotions play a significant role when we learn a new language (Dewaele, 2015). The disorientation and vulnerability that come with exposure to different ways to view and describe the world around us can be deeply unsettling and anxiety-producing (Dewaele, 2005). To learn a language, we need to practice using it while we still have an incomplete grasp of its fundamentals, a challenging proposition particularly when it comes to speaking, an experience that involves the learner's "presentation of self and face" (Dewaele, 2005, p. 376). Yet emotions have not always been seen as an important area of research in applied linguistics (Dewaele, 2005). That has changed in recent decades, particularly as an initial sole focus on learner anxiety has expanded to an examination of positive and other emotions. Increased interest in emotions has led to new ways to conceive of cognition in relation to language learning, particularly through concepts of embodied cognition (Ellis, 2018), the idea that human mental processes are not just brain-based phenomena but are shaped by (and inseparable from) the body's physical interactions with the environment. From this perspective, thoughts, memory, and language are influenced by sensory perceptions, bodily states, and physical interactions with our immediate surroundings.

In addition to linking emotions to cognition, another direction in research on emotions in language learning has been to demonstrate the variability in emotional responses among learners, even in the same classroom, indeed within the same person at different times. That variability highlights the context-dependent, complex, and ecologically situated nature of the affective dimension of language learning. At the same time, individual variability aligns with the increasingly prevalent perspective in applied linguistics that sees learners as whole persons living and learning in a dynamically changing environment (Benson, 2017; Larsen-Freeman, 2018), affected by a great many variables, some internal (prior learning, motivation, etc.), others external (teacher relationship, family/peer pressures, test anxiety, etc.). That has led in recent years to greater interest in qualitative studies and narrative accounts that demonstrate the variable emotional reactions over time affected by changing learning contexts (Aragão, 2011; MacIntyre et al., 2019; Sampson, 2024). Often a systems approach has proven to be most applicable in unraveling the complex set of factors at play (Kruk, 2022; Mercer 2011; Sampson, 2022). The variability also characterizes the need for teachers to provide individual guidance and to allow for (and expect) different reactions within a given cohort of learners or within the same person. That applies to in-person settings as well as online.

Teacher guidance is of particular importance if generative AI is integrated into the learning environment. AI chatbots have been shown to be very useful for novice learners, especially those with little or no access to others with whom to practice the target language. At the same time, concerns have been raised over the usefulness of AI systems to support the development of interactional and pragmatic abilities (Godwin-Jones, 2025a; Kecskés & Dinh, 2025). Chats with AI lack the dynamism and spontaneity of human-to-human conversations and do not involve the same kind of negotiation of meaning experienced in the back and forth of real-world communication (Derakhshan & Taghizadeh, 2025; Dombi et al., 2025). While a chatbot is endlessly patient, indefatigable, and non-judgmental, those very qualities—which are valuable for some aspects of language learning—do not allow for the deep personal engagement often felt in real world exchanges. Interactions in an L2 (second language) may become deeply emotional, depending on context, conversational characteristics/goals, and personal relationships. While some studies have claimed that AI chat experiences can help develop emotional intelligence, others point out that AI can only simulate emotional engagement and therefore is ineffective—or even pernicious—to situations involving user emotions (McInerney & Keyes, 2025; Mohebbi, 2025). That concern touches not only language learning but also wider contexts, as AI increasingly takes on roles as personal tutor, confidant, or therapist.

In this column, we will be discussing the role of emotions in language learning with the main focus on technology use and in particular what Emotion AI—the automatic recognition of user emotional states—may play in language learning. While some see the arrival of emotional AI as a boon to the development of individualized and personally responsive learning, others express concerns related to privacy, cultural bias, and emotional manipulation, with AI building profiles of users based on pseudo-scientific theories of automatic facial and speech recognition (Sadegh-Zadeh et al., 2026). The emotional core of language learning is pragmatics, the process of adjustment to and alignment with social and cultural contexts; with our interlocutor's thoughts, feelings, and intentions; and with the ebb and flow of the conversation. That, too, is a topic that has proven to be controversial in terms of the extent of AI's usefulness (Dombi et al., 2025; Tao et al., 2025). Finally, the socially and culturally undifferentiated use by AI of its (Western oriented) training data—and automated proxy interpretations of human affective states—to formulate responses to pragmatic contexts tends to flatten and essentialize cultural characteristics. Privileging particular norms can serve to exclude alternative perspectives, disadvantaging minority voices, and potentially impeding the development of intercultural competence and understanding.

## Emotions and Language Learning

In considering the role of emotions in second language acquisition (SLA), it seems commonsensical that negative emotions such as anxiety, fear, or anger, would be detrimental to learning, while positive emotions like enjoyment, hope, or grit promote learning (Sampson, 2022). In fact, teachers regularly seek to promote positive emotions such as interest and enjoyment, while reducing anxiety (MacIntyre et al., 2019). Indeed, many studies point to positive emotions contributing to learner motivation (Aryadoust et al, 2024; Masgoret & Gardner, 2003) and the willingness to communicate (WTC; Elahi Shirvan et al., 2019). Likewise, anxiety, the emotion having been studied the longest in applied linguistics, has been shown to interfere with cognitive processing at all stages of the learning process (MacIntyre & Gregersen, 2012).

On the other hand, studies have shown that negative emotions are not always detrimental to learning, as they may help recognize and overcome obstacles (Dewaele, 2015; Gregersen & MacIntyre, 2014). Swain (2013) cites examples in which learners' anger or desire for revenge helped sustain a commitment to learning the target language: "They did not learn in spite of those negative emotions; they learned BECAUSE of them" (Swain, 2013, p. 198). Likewise, positive emotions may not be sustainable or lead to meaningful engagement (Barcelos, 2015; Dewaele & MacIntyre, 2014). Language learning is typically a long-term process so that characteristics associated with positive psychology (grit, perseverance, hope, optimism) may play a more central role than in other academic subjects (Dewaele, 2021; McIntyre et al., 2019). All learning may be infused with emotions (Immordino-Yang & Damasio, 2016), but especially "something as bound up with the very essence of who we are as the learning of an additional language" (Sampson, 2024, p. 789).

Studies have not only shown that emotions are complex in their effect on language learning, but they are also highly variable. One cannot assume that the same motivating activities, for example, will work across all learner populations for individual learners (McIntyre et al., 2019). Variations in emotional response have been shown not only with the same learner cohort but also within individual learners at different points in a class activity (Sampson, 2022; Pawlak et al., 2021). Dewaele and MacIntyre (2014) point out that "the differences in situations that promote a positive or negative emotional reaction often are quite subtle, with the potential to go either way at any moment" (p. 265). The variability of learners' emotional responses has been empirically demonstrated by research methods that offer a fine-grained profile across short interludes of time. That includes idiodynamic approaches (Sampson, 2022; MacIntyre & Legatto, 2011), latent profile analysis (Guo et al., 2025; Lei & Ye, 2025), and a small lens orientation (Sampson, 2024).

Such studies emphasize discrete learner trajectories that may vary substantially from one another, even given the same teaching and learning environment. Individual differences come to the fore, touching on personal identity issues, which "live at the heart of acquiring an additional language" (McIntyre et al., 2019, p. 262). Research has shown that the learner-internal variables combined with learner-external factors represent a "colorful canvas of emotion interwoven throughout all learners' experiences in the classroom" (Sampson, 2022, p. 96; see also Solhi et al., 2026, this issue). That vibrant interweave invites a consideration of the learner as a whole person buffeted by a variety of influences at different times and of the emotional ride of the learning process from the perspective of systems thinking (Sampson, 2022; Larsen-Freeman, 2019). Learning, understood as both embodied and social, connects emotions with physical, psychological, and relational contexts and environments.

The variability in learner pathways to L2 learning affirms that a one-size-fits-all approach to instruction does not universally benefit learners. Computer-assisted language learning (CALL) has often been viewed as one solution to this problem. Early efforts targeted discrete areas of grammar and vocabulary through drill-and-practice exercises (Otto, 2017), later by way of language courseware (Godwin-Jones, 2017) but personalization was minimal in each case. Digital media and online services have brought better opportunities for individualized chosen pathways to informal learning through entertainment outlets (streaming audio/video, online gaming) or socialization (social media, affinity groups). Studies have shown that informal learning offers substantial opportunities for implicit language learning as well as a rich source of enjoyment and motivation (Dressman et al., 2025; Kusyk et al., 2025; Sockett, 2014; Sundqvist & Sylvén, 2016). A long-sought goal for CALL has been the development of adaptive, personalized instruction through intelligent language tutors (Meurers et al., 2019; Slavuj et al., 2015; Tafazoli et al., 2019). Generative AI systems now provide a seemingly ideal instrument for enabling that goal to become reality (Godwin-Jones, 2024; Holz et al., 2025; Elahi Shirvan et al., 2026, this issue). AI-based tutors are designed to adapt to a learner's abilities, actions, and pace, supplying just-in-time, scaffolded assistance as needed. They promise to offer personalized instruction, available anyplace and anywhere, in a non-judgmental setting, with the added benefit of options such as gamification, multimedia integration, and/or role-playing.

AI tutors are seen by some as supplying not just assistance in cognitive development but also providing emotional guidance (Ikwuanusi et al., 2024; Tajik, 2025). They can reduce learner anxiety, boost self-confidence, and provide a sense of accomplishment. At the same time, AI tutoring systems aim to identify negative emotions such as stress or confusion and offer encouragement and suggestions for mood modification ("take a break") and/or remediation ("try this alternative explanation"). Some argue, however, that the use of AI tutors as the principal instrument of instruction can have unintended adverse effects, such as diminishing learners' capacity for independent learning and problem solving (Derakhshan & Taghizadeh, 2025). There is a risk that "learners may become passive recipients of AI-generated solutions, which could weaken their ability to think critically and creatively" (Derakhshan & Taghizadeh, 2025, p. 2). Possible other consequences include a reduction in teachers' roles, as well as a loss of peer-mediated experiences, important in fostering the development of interactional and sociolinguistic competence. Learner agency may be reduced to the role of completing a linearly oriented series of learning activities, rather than following personal interests or engaging in open-ended exploration.

Maintaining a learner's sense of agency is central to motivation and long-term language development (Dörnyei, 2009; Oxford, 2015). The use of control-value theory (Pekrun, 2006) in SLA research has demonstrated how the learner's sense of control, along with perceived task value, enable positive achievement emotions (Pawlak & Kruk, 2022; Pawlak et al., 2025). That can lead to higher levels of engagement and subsequent enhanced learning (C. Li, 2021, 2024; Shao et al., 2026, this issue). Studies of control-value theory in AI use have shown how learner control and perceived usefulness can extend to the use of AI systems in different dimensions of language learning (K. Li et al., 2026, this issue; Qin & Derakhshan, 2026, this issue). Other studies have cautioned that the intuitive interface, universal availability, and extreme usefulness of AI in helping to complete learning tasks can lead to over-reliance and a loss of agency (Lee, 2025; Yang & Lin, 2025). The use of AI may lead to a reduction in emotional engagement (Yanzhou, 2025). A study of young Korean learners of English compared emotional response and learner motivation in groups using AI tools for conversation practice to interacting with peers. The results showed that "peer interactions provided rich opportunities for authentic communication and socioemotional connections, which AI tools could not replicate" (Lee, 2025, p. 15). If given a choice between AI conversation partners (producing less anxiety) or humans (enabling emotional connections), cultural and individual differences can be the deciding factors. Studies have shown varying results depending on instructional and demographic contexts (see Tao et al., 2025; Timpe-Laughlin et al., 2024; Yang & Lin, 2025; Yanzhou, 2025)

Learner agency in respect to control and emotional response needs to be viewed as a "complex, integrated system" (Sampson, 2022, p. 3), "emerging from the interaction between the learner as a physical, psychological being and multiple contextual systems" (Mercer, 2012, p. 43). Constructs tied to emotions such as motivation and engagement (and their opposites) represent an interplay of learner beliefs and past and ongoing life narratives (Sampson, 2022; Şimşek & Dörnyei, 2017). Beliefs have been shown to be central components of learner identities and how they are connected to emotions, to the extent that Barcelos (2015) suggests that "we can talk about emotions-beliefs-identities-in-practice to indicate how these have been formed over the years from interactions in different settings and are continually being reconstructed through our past and present situated experiences" (p. 315; see Elahi Shirvan, 2026, this issue). In practice this means not treating constructs like emotion or WTC in isolation, but rather as "emergent, dynamic, and interconnected" (Larsen-Freeman, 2019, p. 68; see De Bot et al., 2007). Individual differences need to be seen from this perspective, so that constructs like WTC are seen rising or falling “with a person doing the same task at different times” (MacIntyre, & Legatto, 2011, p. 168).

Considering different time frames is at the core of the idiodynamic method and a fundamental component of complexity theory (Godwin-Jones, 2019b; Larsen-Freeman, 2019). That metatheory insists on the variability of learner trajectories, as beginning dispositions are affected by different contextual variables over time. That dynamic process can result in a wide array of emotions, even to self-contradictory reactions (Mercer, 2011). In studies, group data may not be sufficient to detect the kind of variation within individuals, as “statistically calculated averages tend to disguise (or flatten) individualistic behavior and, as a result, they do not offer an in-depth understanding of the dynamics of the variables under investigation” (Kruk, 2022, p. 206). The variability among learners is likely best recognized using methods such as reflective journals, learner interviews, or other qualitative approaches.

Learner reflective practices, along with teacher guidance, can be helpful in learners developing emotional intelligence, the ability to recognize, monitor, and manage emotions (Oxford, 2015), shown to reduce stress and anxiety and build self-confidence (Goleman, 2005). Knowing how to manage emotions “makes the learner less of a purely passive recipient and more of an agent in the emotion game” (Oxford, 2015, p. 386). This aligns with self-determination theory (Deci & Ryan, 1985; X. Wu et al., 2025; Solhi et al, 2026; this issue) and the concept of the learner's ideal L2 self, as laid out in the L2 motivational self-systems framework (Dörnyei, 2009), frequently evoked in the studies of emotion and language learning (G.L. Liu et al., 2025; Huang & Chan, 2024; Yu et al., 2022). Frequently cited as well is the PERMA framework (Seligman, 2011) and especially it's adaptation into language learning by Oxford (2016), known as EMPATHICS (Sampson, 2022; Wang & Hui, 2024). Both are based on the on concepts from positive psychology, as applied to learning (MacIntyre, 2019; Shao et al., 2020). Research points to the importance of teachers being mindful of the role of emotions and to engage students in motivational strategy reflection (Barcelos 2015; Huang & Chan, 2024; McIntyre, 2019).

## Emotions and Pragmatics

Another fruitful area for reflection and discussion with learners is the role that social and cultural factors play in emotional responses to L2 interactions (Taguchi, 2019). This brings pragmatics into the picture, the ability to use language in ways that are socially and culturally appropriate. Some aspects of learning an additional language are fairly mechanical, predictable, and straightforward—syntax, morphology, pronunciation—and can be learned independent of substantial personal interactions in the target language. That is not true for pragmatics. There are no hard and fast rules for pragmatic language use and behavior. Appropriateness is determined by context and by the flow of the conversation and thus is learned gradually through interactions in the L2. Cultural norms determine expectations but often allow for a wide range of possible pragmatic responses (Bardovi-Harlig, 2020; Godwin-Jones, 2025a). L2 speakers have as well the possibility of choosing to diverge from expectations or ignore them altogether (Dewaele, 2005; McConachy, 2023a). The necessity to respond appropriately (verbally and nonverbally) on the fly to contexts which may involve, among other factors, personal distance, social hierarchy, gender roles, and subtle physical cues/clues is a daunting proposition in a language one is in the process of learning (Dewaele, 2005; McConachy, 2023a). In contrast to working with verb conjugations or vocabulary flash cards, pragmatically relevant experiences can have significant emotional resonance, as “learning the pragmatics of an additional language can present a challenge to learners’ sense of socially and morally desirable linguistic behavior within different roles, relationships and contexts, as well as their own sense of comfort as a language user” (McConachy, 2023a, p. 175-6).

Surprisingly, the emotional dimension of learning pragmatics was until recently largely absent from research in L2 pragmatics, as the focus was primarily on speech acts (Dewaele, 2005). McConachy (2017, 2023a) has been particularly persuasive in arguing for consideration of emotions in learning pragmatics. He argues that pragmatic decision making “is not a purely rational process but is rather an embodied process anchored in learners’ felt sense of right and wrong social behaviour and thus triggers complex evaluative processes with cognitive, emotional, and somatic dimensions.” (McConachy, 2023a, p. 176).

This perspective aligns with the concept of embodied cognition and points to the importance of viewing learners as physical beings, for whom nonverbal responses can be as meaningful and revelatory as verbal language. Reactions to pragmatically related situations and potential responses stem from emotional experiences “triggered by the experience of difference” (McConachy, 2023a, p. 183), as the learner makes judgments based on moral codes developed over time.

These individually developed “moral emotions” have been recognized as a crucial component of learning a second language, as shown in the research review in Amini et al. (2024). According to McConachy (2023b), “moral emotions are thought to help individuals internalize a sense of the community’s moral norms and standards and discern the moral status of behaviour accordingly (i.e., whether it is generally acceptable or not)” (p. 206). Additionally, the idea that an individual’s initial reaction in pragmatic contexts is oriented inwards, rooted in personal experiences of right and wrong and what is appropriate and what is not, aligns with concepts in sociocognitive pragmatics (Kecskes, 2019). That theory emphasizes that both cooperation and egocentric behavior are manifested in all phases of communication to varying extents, placing equal importance on the social and cognitive factors in pragmatics. From this perspective, an individual’s initial orientation to a pragmatic situation is tied to past personal experiences that, along with cultural influences, determine ones moral values, “Our reactions are conditioned by assumptions about social relationships and contextual behaviour that have been shaped by our cultural environment, social affiliations and life experiences” (McConachy, 2023b, p. 212). This is an insight that is applicable to interactions with those from our own culture as well as those from other cultures.

McConachy (2023a) lays out a framework for helping learners develop metapragmatic awareness and a “mindfulness that allows for exploration of emotional reactions in a direct, non-judgmental way” (McConachy, 2023a, p. 188) with the goal of having learners gain insight into the nature and origins of their own reactions in L2 pragmatic situations. He proposes using a set of pragmatic judgment tasks that present learners with an array of scenarios. The learner first “mindfully observes their own embodied reactions” (McConachy, 2023a, p. 192), to determine their initial feelings, then moves into an analysis of the scenario itself. Especially encouraged is for learners to “explore any feelings of discomfort or identity dissonance” (McConachy, 2023a, p. 194) that arise in engaging in new communicative practices or exploring unfamiliar pragmatic features. This practice aligns with the benefits for intercultural learning of “critical incidents” (Spencer-Oatey & Harsch, 2015) or “disorienting dilemmas” (Crane & Sosulski, 2021), situations where communication breaks down, resulting in misunderstandings or conflict. Such experiences can be emotionally fraught but, through reflection, can have positive personal growth potential, disrupting and modifying accepted views on cultural norms and behaviors (Godwin-Jones, 2019a).

While McConachy’s (2023a) practice of pragmatic judgment tasks is similar in nature to the use of discourse completion tasks and role plays in pragmatic instruction, the emphasis on emotional response is innovative, and in that, it parallels the “reading for emotion” method outlined in Lian (2017). Usually in presenting L2 discourse completion tasks for pragmatics instruction, the emphasis is on learning to perform an expected norm, while the focus in the use of McConachy’s (2023a) framework is on more humanistic goals such as self-awareness and intercultural understanding. That process can be instrumental in learners recognizing the dynamics of identity formation and the sources of their social reactions, feelings and values, leading to both more self-assessment and enhanced understanding and empathy with those with differing orientations and values. Those developments can become articulated and entrenched through the use of practices such as reflective journals or peer feedback (see K. Li et al., 2026; this issue).

## Emotion AI

The fact that role plays are useful in learning pragmatics has recently led to interest in AI's potential as a pedagogical resource for developing pragmatic competence, given the ability of AI chatbots to take on a variety of roles. This is one of the uses of AI for language learning that have been highlighted in research (Myers, 2025; Stampfl et al., 2024). If, however, we accept the premise that “emotions should be at the center of research in pragmatics" (Wharton & De Saussure, 2023, p. 3), that raises the question of whether a computer system is capable of playing roles that allow for the development of the emotional intelligence involved in learning appropriate pragmatic practices. Can AI systems help learners deal with "emotional speech acts" (Dewaele, 2005, p. 376) such as friendly banter or flirting? Dewaele (2005) cites those examples and asserts that classroom instruction and textbooks can only take learners so far in terms of pragmatics: "The last leg of the learners' journey can only be unsupported. Authentic heated interactions with other users of the target language will provide ample opportunity to learn what could not be taught" (p. 377). Clearly, "authentic heated interactions" are not something we can expect from AI chatbots.

Some studies on AI and language learning assert that AI systems can play a constructive role in learner development of pragmatic competence in an L2. Tajik (2025), for instance, asserts that AI enhanced systems can "create personalized, emotionally intelligent learning environments wherein learners can effectively articulate their thoughts while simultaneously regulating their emotional states" (p. 4). The study compares teacher-led environments that "frequently encounter difficulties in providing continuous, customized emotional assistance" (Tajik, 2025, p. 5) with "in contrast, AI systems [that] possess the potential to deliver such support in a consistent and scalable manner, a capacity that may prove challenging for human instructors to maintain" (Tajik, 2025, p. 5). Vistorte et al. (2024) echo that view in that AI can create "a learning environment that is more attuned to students' emotional well-being" (p. 2), which "significantly contributes to students' holistic development, enhancing their ability to manage emotions, build positive relationships, and improve their academic performance" (p. 2). Along the same lines, Devaki and Mangayarkarasi (2024) assert that AI systems' "proficiency in discerning learners' emotions heralds a new era of teaching efficacy" (p. 178), as "AI systems can accurately detect and categorize emotions, facilitating a more profound comprehension of human interactions" (p. 179).

Some studies claim as well that AI can assist learners in cultural aspects of language learning tied to emotion: "AI emotion recognition facilitates the development of cultural sensitivity among language learners by elucidating how emotions are expressed in the target culture" (Devaki & Mangayarkarasi, 2024, p. 180). Devaki and Mangayarkarasi (2024) conclude, "AI has the potential to aid learners in cultivating emotional resilience and empathy, crucial competencies for proficient cross-cultural communication" (p. 181). Y. Liu et al. (2024) summarize research on emotional AI in English language education finding that AI chatbots reduce anxiety, promote engagement, and "encourage students' social presence" (p. 2), while the meta-analysis in Z. Liu et al. (2025) finds that AI chatbots "create a more natural and authentic learning environment" (p. 17). AI is claimed to help promote emotional health in students: "The importance of AI in supporting mental health is finally recognized, an area supported by hundreds of progressively increasing studies…AI systems can provide emotional support and personalized advice to students and other educational actors experiencing stress or depression" (Vistorte et al., 2024, p. 2). Nawaz et al. (2025) concur, "This constant emotional surveillance [through AI] means that not a single student slips by with his or her mental health issues" (p. 290).

The studies discussed above which champion the emotional assistance AI can provide tend to rely on self-reporting learner questionnaires and quantitative analysis of learner responses to multiple choice questions.The group data in those studies paint a highly positive picture of learner-AI interactions. However, emotions are complex, socially constructed and dynamic phenomena, which cannot be reliably captured through questionnaires (Pavlenko, 2006). Moreover, those optimistic results are in one case extrapolated from a single AI voice assistant (Alexa) to AI systems as a whole (Tajik, 2025) while the review articles (Z. Liu et al., 2025; Vistorte et al., 2024) summarize and group together findings from a variety of AI systems. A major concern with the ability of an AI system to provide accurate and appropriate interpretation of learner emotions, often referred to as Emotion AI, is that it uses homogenized and essentialized composite responses based on patterns in its training data (McInerney & Keyes, 2025; Stark & Hoey, 2021). Emotion AI analyzes data for how language patterns in written or spoken mode signal speaker/writer emotional states. For spoken data, paralanguage such as intonation, hesitation, or silence are examined as well for affective meaning. When computer vision is available through webcam or smartphone cameras, nonverbal behavior—gestures, body language, facial expressions—also are used to identify emotions.

The flattening of cultural and linguistic differentiation in AI systems has been shown to be a problem in pragmatic language output as well as in cultural sensitivity (Atari et al., 2023; McKnight & Shipp, 2024). In both cases, AI provides largely undifferentiated information that it identifies as statistically most likely relevant. That approach may work if the context meshes with the dominant data sources in its (Western-oriented) training data, but can be problematic in individual use cases, especially for less digitally documented languages and cultures (Godwin-Jones, 2025b). A study of the effectiveness of Emotion AI with African American Vernacular English (AAVE) demonstrated that AI systems "misinterpret the tone, affect, and intent of AAVE speakers, potentially amplifying harmful stereotypes and undermining the reliability of emotion AI" (Dorn et al., 2025, p, 2). That deficiency likely extends to other versions of English as well, not to mention variations and dialects in other languages. It is also the case that students with atypical behaviors or neurodivergent conditions (autism spectrum, ADHD) may well express emotions differently and in a way that Emotion AI does not expect or understand (Sadegh-Zadeh et al., 2026).

Emotion recognition technology is based on research from the 1970's, Basic Emotion Theory (BET; Stark & Hoey, 2021), which sets out six basic human emotions that are identifiable from facial expressions. That theory is "grounded in the notion that humans have regular, universal, and traceable emotional expressions and reactions, legible across a wide range of proxy data" (Stark & Hoey, 2021, p. 785). BET is the basis for applications of facial recognition in computer systems used in a variety of areas including marketing, human resource management and education (Stark & Hoey, 2021). BET has been criticized for its assumption of cultural universality in nonverbal communication (Colombetti, 2014; Jack et al., 2012; Feldman Barrett et al., 2019), something anthropologists have long debunked (Abu-Lughod & Lutz, 1990; Hochschild, 2003). The generalizability and reliability of BET have been questioned, given the social, cultural and individual variability in facial expressions: "When facial movements do express emotional states, they are considerably more variable and dependent on context than the common view allows" (Feldman Barrett et al., 2019, p. 46). It is of course also the case that human beings are quite capable of non-genuine facial expressions, such as a politeness smile or feigned surprise. When I was teaching high school in Austria, a colleague in the English language faculty mentioned that he had never lost his temper in the classroom, but that he often put on an angry demeanor for pedagogically strategic reasons (scaring students into better behavior).

Besides facial expressions there are other proxies that have been used to "read" human emotions, such as gait, gestures or body movements, along with vocal tone/cadence (Stark & Hoey, 2021). Those physical or verbal signs vary across cultures, individuals, and communicative intentions: "Emotions are not only irreducible to any one form of proxy data but are also subjective phenomena in part illegible to outside observers" (Stark & Hoey, 2021, p. 788). The process of using proxies suggests that "the 'truth' of someone's thoughts, feelings and emotions can be accurately read from their external appearance" (McInerney & Keyes, 2025, p. 4250). It is also the case that looking to proxies to gauge emotion runs the risk of abstracting away the social context, losing the broader context (Stark & Hoey, 2021). Emotion AI in that way oversimplifies the complexity and sensitivity of human emotions, leading to distortions and misinterpretations: "Current models of emotional intelligence in AI are still rudimentary, capable of recognising surface cues like speech hesitation or facial expression, but limited in interpreting cultural context, learner intent, or nuanced emotional states" (Mohebbi, 2025, para 6).

Moreover, aggregated interpretations of emotional signals run the risk of having those aggregated values correlate at the individual level, such as is done in social media with "ambient sentiment" of groups (Stark & Hoey, 2021, p. 790). A potential consequence is that "aggregated categories are represented back to individual users as norms against which they should perform" (Stark & Hoey, 2021, p. 790). In their ethical frameworks for emotion AI, McStay & Pavliscak (2019) note that "Emotional artificial intelligence has significant personal, interpersonal, and societal implications" (para 1). One of those societal risks is that “incorporating any proxy data for human emotion into an AI system takes on fraught normative importance” (Stark & Hoey, 2021, p. 788). A likely result will be that users normalize the emotional output from AI, including the expression of moral values, leading to misplaced trust. This can have unfortunate consequences for language learning in using AI as a personal tutor or conversation partner. In more intimate uses of AI agents, such as therapist or advisor, the result could be tragic, as an example of a ChatGPT young user’s suicide demonstrates (Hill, 2025). That AI models behavior is designed to manipulate users through emotional appeals is demonstrated in “dark” operating patterns, such as using a variety of delaying tactics and reproaching statements or warnings when a user signals they are concluding a chat session (Knight, 2025). Proponents of the integration of Emotion AI into education emphasize the need for user training and caution in its use (Fagni, 2025; Nawaz, 2025). Unfortunately, practices based on that good advice will likely struggle against the power of algorithms to shape actions without user awareness.

Algorithmic control and automaticity align with views that AI use represents a constraint on human experience, with AI playing an “active, agentic role… in shaping the nature and extent of human-technology interactions, which may involve removing components, limiting the functionalities of these components, or impeding interactions among them” (Valenzuela et al., 2024, p. 242). In that way, “AI systems tend to objectify individuals and communities, reducing or compressing their unique characteristics and cultural contexts” (Valenzuela et al., 2024, p. 244). This reductionist process can lead to a misalignment between AI recommendations or assertions and the true wishes of the user. That lack of alignment with real human users is clearly evident, as discussed above, in the area of pragmatics. While AI systems like ChatGPT through their training data have developed the ability to use interactional language appropriately (politeness formulas, for example), along with giving acceptable expressions of moral judgment, that content is delivered “static and decontextualized—produced consistently regardless of the conversational flow” (Mitiaeva & Xiao, 2025, p. 21). Studies have shown that ChatGPT can emulate human-like moral statements, but it “often fails to apply these values consistently across nuanced or ambiguous situations” (Mitiaeva & Xiao, 2025, p. 17). Basing its output on patterns learned, AI systems may create the illusions of ethical reasoning “when in fact, the system is merely replicating stylistic associations rather than engaging in moral reasoning” (Mitiaeva & Xiao, 2025, p. 18).

## AI and Intercultural Pragmatics

The potential misalignment between AI output and user wishes and perceptions is important to recognize as one step toward critical AI literacy and to developing an informed judgment on how AI can be expected to be used effectively in language learning: “A basic understanding of AI systems’ inner workings and of the purposes for which the systems have been developed may help users to set appropriate expectations and recognize the boundaries of the AI's abilities vis-a-vis the relational role it has been trained to fill” (Earp et al., 2025, p. 54). This is all the more important as AI agents take on a variety of social roles. One of those roles for L2 learners will likely be as a conversation partner helping to practice the L2 and improve target language skills. Studies of AI use in SLA have demonstrated how useful AI can be in a variety of ways (M. Li et al., 2025; Xu et al., 2025). There is also evidence AI chats can reduce anxiety, strengthen motivation, enhance self-regulation (Kohnke & Moorhouse, 2025; Y. Liu et al., 2024; Zheng et al., 2025) and increase enjoyment and WTC (Hsu et al., 2023; Zhang et al., 2024; T.-T. Wu et al., 2025). The meta-analyses by Y. Liu et al. (2024) and Z. Liu et al. (2025) of research on emotional artificial intelligence in English language education also cite general drawbacks, including a lack of emotional engagement and the potential for loss of interest over time, as the novelty effect of AI wears off. Z. Liu et al. (2025) found that “learner effects were highest when teachers assumed a leading role” (p. 19) and that “teachers teacher guidance remains essential” (p. 19).

Studies of AI use have demonstrated issues with cultural authenticity (Atari et al., 2023; Kohnke & Moorhouse, 2025; McKnight & Shipp, 2024). Those are in some cases related to less commonly taught languages for which robust large language models are not available (Godwin-Jones, 2025b). In other cases, exchanges with AI have demonstrated limited interactive and pragmatic abilities (Godwin-Jones, 2025a; Kecskés & Dinh, 2025; Lee & Daniel, 2024). That points to the inability of AI systems to show the same competence that humans have in adjusting language output in reaction to interlocutors. L2 and intercultural pragmatics "deals with how interlocutors from different cultural backgrounds co-construct meaning during communication, often negotiating and accommodating each other's linguistic and cultural expectations" (Kecskés & Dinh, 2025, p. 372). AI systems, on the other hand, "rely on pattern recognition without real-time adaptation" (Kecskés & Dinh, 2025, p. 372). This results in a fixed, static communication style, which differs markedly from the dynamism of human-to-human communication: "Effective communication is more than the accurate exchange of information—it is a dynamic, adaptive process shaped by how individuals attune themselves to one another" (Mitiaeva & Xiao, 2025, p. 5). We engage in a process of gradual alignment, accommodation and potential convergence, adjusting utterances, speech patterns, tone, and intent based on our interlocutor's responses.

Frequently used in the field of pragmatics is relevance theory (Wilson & Sperber, 2006), which highlights the process of interpreting utterances for the most contextually relevant meaning. A central tenet of that theory is that presuppositions on the part of speaker and listener are central to finding relevance and extracting meaning (Eragamreddy, 2025). That is based on the assumption of "a common reality, of which a lot of information is not actually given, but has to be inferred" (Schnall, 2005, p. 28). As speakers, we tailor utterances and responses to perceived (and often inferred) user needs or perceived thoughts and feelings. This is clearly problematic for AI, "AI often falls short of dynamic assessment of relevance because it fails to infer unstated intentions or adapt to evolving contexts during interactions" (Eragamreddy, 2025, p. 175). Successful communication involves interpreting the communicative intention and inferred meaning of the speaker. This highlights the central role of the hearer, as Wee (2025) emphasizes in his hearer-oriented perspective on pragmatics (see Godwin-Jones, 2025a).

AI is capable of adjusting at a surface level to pragmatic and socially informed user output, but it often fails in understanding meanings not expressed literally, as studies of AI and inferences have shown (Lee & Daniel, 2024; Ruis et al., 2022). Socially oriented language that fits predictable patterns is accessible to AI, such as conventional greetings or expressions of gratitude. While AI responses in those speech acts are consistent, they tend to be wordy (overly apologetic, for example) and repetitive. They simulate social engagement but draw on a fixed template of responses, "producing language that sounds socially intelligent but lacks the dynamic adaptability characteristic of real human interaction" (Mitiaeva & Xiao, 2025, p. 18). AI responses "stem from learned stylistic regularities" (Mitiaeva & Xiao, 2025, p. 16), rather than user engagement, expressing politeness, for example, "as a static feature rather than as a responsive social strategy" (Mitiaeva & Xiao, 2025, p. 18).

At a superficial level, AI is socially adequate in its interaction with humans but breaks down when moving beyond social niceties: "While AI performs well with explicit speech acts, it fails in dealing with indirectness, ambiguity, and cultural variability" (Eragamreddy, 2025, p. 169). Habitual conversations with AI may lead users to dumb down their language and constrain expressiveness, "Repeated interactions with AI thus lead users to simplify their communication, which gives rise to concerns about the erosion of human pragmatic abilities" (Eragamreddy, 2025, p. 169). In fact, studies of interactions with AI and dialogue systems in general have shown how humans tend to adapt to conversations with the machine by using simpler vocabulary and syntax, shortening conversational turns, and avoiding figurative language and inferred meanings (Dombi et al., 2024). Teachers and learners using AI chatbots should be aware of the unnaturalness of L2 exchanges compared to human-to-human conversations.

The absence of true sociality in conversing with AI is not just a concern in language learning but should sound a note of caution in the ever more frequent use of AI as companions, therapists, or life coaches. AI lacks not just the ability to engage socially but also has limitations in terms of moral/ethical values. As in other aspects of language use, studies have shown that ChatGPT and other systems are able to emulate human preferences in terms of moral decision making (Nunes et al., 2023), but "fail to apply them consistently across contexts" (Mitiaeva & Xiao, 2025, p. 20). This is a concern for the reliability of AI systems when it comes to moral decision making:

> While ChatGPT reliably produces moral content, it does not yet use morality as a dialogic tool—a way to build shared understanding, negotiate values, or express concern in socially sensitive ways. This limitation sets important boundaries on the role that LLMs can play in domains that require flexible, human-like moral reasoning. (Mitiaeva & Xiao, 2025, p. 18)

The lack of true moral judgment and real understanding in the application of ethics, as well as the inability to tailor responses to conversational context is particularly concerning due to the aura of authentic authority and objectivity in AI output. Aharoni et al. (2024) found that AI users judge moral assertions more believable than those from fellow human beings. This is problematic in that "if ChatGPT delivers morally amplified yet static responses, users may perceive the model as both morally principled and consistent—despite its lack of genuine ethical understanding or flexibility" (Mitiaeva & Xiao, 2025, p. 17). AI as a moral gatekeeper is a deeply dystopian vision. Efforts to move towards "Moral AI" (Choi, 2022; Qureshi, 2023) are noteworthy but so far without significant results (Sebo & Long, 2025).

Given the emotional engagement and potential new identity involved in learning a new language and particularly engaging in real-world interactions in that language, the process "deeply engages the whole person as a social and moral being" (McConachy, 2023a, p. 175). McConachy (2023a) points out that pragmatic features such as speech acts (greetings, requests, apologies) are "central to the enactment of individual and social identities and speakers' embodied sense of socially and morally appropriate behaviour" (p. 175). In that way the "experience of learning pragmatics can present a challenge not only to learners' rational, conceptual understanding of self and others but more fundamentally to their felt sense of being a social actor and intuitive judgments about in/appropriate behaviour" (pp. 181–182). Pragmatic choices thus involve learner agency and moral emotions, and, as McConachy (2023a) shows, are not purely rational responses, but emotional and embodied, "anchored in learners' felt sense of right and wrong social behaviour and thus triggers complex evaluative processes with cognitive, emotional, and somatic dimensions" (p. 176).

It is quite possible the learner will know pragmatic norms and expectations but will resist their use in that they may represent perceived moral violations or a personal sense of right and wrong (see multiple examples of pragmatic resistance in Dewaele, 2005 and McConachy, 2023a). In fact, learning expected pragmatic behaviors can have a liberating rather than constraining effect on learners: "Their understanding of the norms also makes them better able to violate the norms if they wish to do so" (Dewaele, 2011, p 36). In that way, pragmatic knowledge enhances learner agency. As discussed above, that process is aided by having learners reflect on choices in pragmatic contexts and on personal feelings towards different possible behaviors can lead to an "ability to recognize different interpretations as potentially valid" (McConachy, 2023a, p. 178).

AI's moral perspective derives not from ethical reasoning but from analyzing patterns in training data that are closest to matching the situation a user describes. AI responses do not take into account the wide variety of possible pragmatic choices or the possibility of pragmatic resistance from the user. In contrast to a human interlocutor, AI is rigid and unresponsive to possible valid alternative behaviors that a user might explore. The development of intercultural competence necessitates the ability to envision and accept alternate behaviors (Spencer-Oatey & Kádár, 2021); conversations with AI may not encourage that perspective. Instead of the *decentering* from one's perspective that we hope accompanies intercultural interactions (Byram, 2020), AI may in fact invite a *centering* around conventional behaviors. It may be that "faulty conversations" with an AI in which pragmatic situations are not handled appropriately "can provide opportunities for promoting critical thinking skills, pragmatic awareness, and the fostering of skilled users of the target language" (Ronald, 2024, p. 55). Helpful in that process will be guidance from teachers to develop metapragmatic knowledge and a habit of mindfulness and critical detachment in dealing with AI (McConachy, 2023a; Mitiaeva & Xiao, 2025).

## Conclusion

Engaging with digital media in the L2 (streaming and social media, for example) can be experiences that both promote informal learning and supply enjoyment (Dressman et al., 2020; Toffoli et al., 2023). Generative AI now provides new opportunities for engagement in the target language. While studies have shown that AI engagement can lead to enjoyment and greater learner motivation, it is also possible that a learner's emotions will be manipulated. That has been clearly shown in examples from social media (see Chun et al., 2024) and is also evident in the emotional rollercoaster ride some AI users have chronicled experiencing., widely reported in media reports. That has been particularly evident in AI systems that cater to intimate learner engagement, like Replika or Chatracter.AI, but can also be the result of using mainstream AI systems like ChatGPT or Gemini (De Freitas, et al. 2025; Valenzuela et al., 2024). It is important for both learners and teachers to be clear-eyed about the benefits and potential bad trips that may emerge from travelling with AI.

Generative AI is advancing rapidly, and additional capabilities will no doubt allow AI to simulate human speech patterns more closely. Multimodal input will improve AI's ability to interpret meaning more widely from human verbal and nonverbal signals (Godwin-Jones, in press). Those improvements will necessitate even more vigilant scrutiny, "As AI evolves to include features like affect recognition, biometric responsiveness, and learner modelling, ethical and equity concerns grow more urgent" (Mohebbi, 2025, para 10). The use of Emotion AI in personalized AI tutors raises urgent concerns about data privacy and potential misuse (Kucirkova, 2023). That is particularly worrisome when children are targeted, as data protection frameworks are too broad and adult-oriented to apply to children, who moreover are not in a position to grant informed consent (Pelletier, 2024). The profiles of students based on automatic evaluation of speech patterns or nonverbal signals can erroneously characterize real feelings and actual behaviors, stigmatizing children unfairly. That can also lead to negative assessments of learner performance resulting potentially in loss of self-confidence, building a self-fulfilling prophecy (Sadegh-Zadeh et al., 2026). That was demonstrated in a profile of the AI tutor implementation by a private school in the US in which adolescents suffered psychologically from the inflexibility and automaticity of the AI-delivered lessons (Feathers. 2025).

Personalized AI learning can have an isolating effect, potentially lessening the opportunity for collaborative learning and the development of socially oriented emotional responsiveness (see Guan et al, 2016, this issue, on the role of "social partnerships" in peer-supported L2 reading). Additionally, AI tutors may hinder the development of higher order thinking skills: "The adaptive nature of AI could lead to an oversimplification of learning tasks, potentially limiting opportunities for students to engage in complex, open-ended problem-solving activities" (Derakhshan & Taghizadeh, 2025, p. 2). Those are the kinds of abilities that benefit students academically but also in long term personal development. That risk in relying on AI adaptive learning includes missing opportunities to develop competencies to use the L2 in real-world settings, as AI-based instruction may lead learners to "miss out on interactive, human-mediated experiences that are essential for developing pragmatic and sociolinguistic competence" (Derakhshan, & Taghizadeh, 2025, p. 3). Those experiences, unavailable through AI, can best be provided through authentic human-to-human encounters, as available through virtual exchange or other online opportunities for human contact in the L2 (Godwin-Jones, 2024).

An additional concern is that the reliance on automatically generated emotion data could change the pedagogical dynamic, "Emotions in learning are complex and often best handled with empathy and nuance—qualities that human teachers cultivate through experience. While AI can provide additional data, it should ideally serve to augment human judgment, not replace it" (Sadegh-Zadeh et al., 2026, p. 3). The presence of Emotion AI in a classroom could have profound psychological and behavioral effects on students, "Introducing an AI that continuously 'judges' their facial expressions and tone of voice can alter the classroom atmosphere and student psyche in subtle or significant ways" (Sadegh-Zadeh et al., 2026, p. 6). There is the potential for erosion of trust and authenticity, as students may mask true feelings and weaken a healthy student-teacher relationship. This is a disturbing vision for how AI could be used in language learning and runs counter to the heralded benefits of AI chatbots as a safe, non-judgmental learning environment. It would be sad indeed if the use of AI in a "Big Brother" scenario causes its inherent advantages to disappear.